\documentclass[aps,prx,reprint,superscriptaddress]{revtex4-1}
\usepackage{color}
\usepackage{soul}
\usepackage{dcolumn}
\usepackage{graphicx}
\usepackage{amsmath}
\usepackage{amssymb}
\usepackage{bm}

\begin{document}
\mbox{}



\title{Quantum electrodynamics of a superconductor-insulator phase transition}




\author{R. Kuzmin}
\author{R. Mencia}
\author{N. Grabon}
\author{N. Mehta}
\author{Y.-H. Lin}
\author{V. E. Manucharyan}

\email[]{manuchar@umd.edu}

\affiliation{Department of Physics, Joint Quantum Institute, and Center for Nanophysics and Advanced Materials,
University of Maryland, College Park, Maryland 20742, USA.}



\date{\today}

\maketitle

\textbf{A chain of Josephson junctions implements one of the simplest many-body models undergoing a superconductor-insulator (SI) quantum phase transition between states with zero and infinite resistance~\cite{sachdev2007quantum, bradley1984quantum}. Apart from zero resistance, the superconducting state is necessarily accompanied by a sound-like mode due to collective oscillations of the phase of the complex-valued order parameter~\cite{fazio1996tunneling, basko2013disordered}. Exciting this phase mode results in transverse photons propagating along the chain. Surprisingly little is known about the fate of this mode upon entering the insulating state, where the order parameter's amplitude remains non-zero, but the phase ordering is ``melted" by quantum fluctuations~\cite{sondhi1997continuous}. Here we report momentum-resolved radio-frequency spectroscopy of collective modes in nanofabricated chains of Al/AlOx/Al tunnel junctions. We find that the phase mode survives remarkably far into the insulating regime, such that $\textrm{M}\Omega$-resistance chains carry $\textrm{GHz}$-frequency alternating currents as nearly ideal superconductors. The insulator reveals itself through broadening and random frequency shifts of collective mode resonances, originated from intrinsic interactions.
By pushing the chain parameters deeper into the insulating state, we achieved propagation with the speed of light down to $8\times 10^5~\textrm{m/s}$ and the wave impedance up to $23~\textrm{k}\Omega$. The latter quantity exceeds the predicted critical impedance by an order of magnitude~\cite{PhysRevB.37.325, korshunov1989effect, Choi1998Quantum, giamarchi2004quantum, bard2017superconductor, cedergren2017insulating}, which opens the problem of quantum electrodynamics of a Bose glass insulator for both theory and experiment~\cite{wu2018theory, bard2018decay, houzet2019microwave}. Notably, the effective fine structure constant of such a 1D vacuum exceeds a unity, promising transformative applications to quantum science and technology.}

\begin{figure*}
	\centering
	\includegraphics[width=\linewidth]{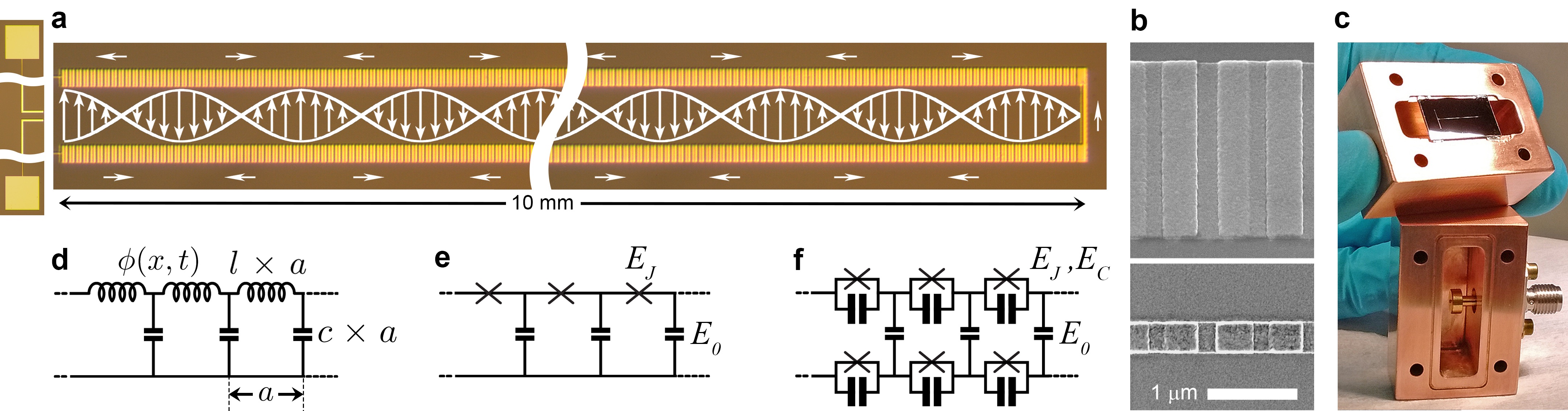}
	\caption{(a) Optical photograph of the Josephson transmission line with a dipole antenna at the left end and a short-circuit termination at the right one. The phase mode corresponds to AC current through the junctions and AC electric field across the two chains. (b) Scanning electron microscope images of chain junctions. (c) Photograph of a chip mounted in a single-port copper waveguide. (d) Linear circuit model for the propagating phase mode at long wavelength (see text). (e) Minimal model of a Josephson chain exhibiting a BKT transition.  (f) Minimal circuit model of the double-chain device from (a). Here the capacitance $c$ is due to electrostatic coupling between the chains and every superconducting island has a random charge offset.}
	
	\label{fig:Fig1}
\end{figure*}

Our devices consist of two closely spaced parallel chains of over 33,000 junctions fabricated on an insulating silicon chip~(Fig. \ref{fig:Fig1}a,b). The chains are short-circuited at one end and connected to a dipole antenna at the other end for coupling to external signals. The chip is suspended in the center of a metallic waveguide box with a single broadband microwave input/output port~(Fig. \ref{fig:Fig1}c). This wireless interface minimizes coupling to stray modes and allows collection of all the energy radiated off-chip. In the superconducting state, the device can be viewed as a telegraph transmission line~(Fig. \ref{fig:Fig1}d) defined by the capacitance $c$ between the chains and the Josephson inductance $l$ per unit length. The capacitance gives inertia to the phase field $\phi(x,t)$ and its value is approximately given by the vacuum permittivity $\epsilon_0$ adjusted by the dielectric constant of silicon. The inductance $l$ is the inverse phase-stiffness of the Cooper pair condensate and it can largely exceed the vacuum permeability $\mu_0$. This effect is a low-dimensional analog of Meissner's diamagnetism. Therefore, a sound-like wave associated with the small-amplitude oscillations of $\phi(x,t)$ is a hallmark of superconducting order in our 1D system. Known as the collective phase mode, this wave is equivalent to transverse one-dimensional photons with a velocity $v = 1/\sqrt{lc}$ and a wave impedance $Z = \sqrt{l/c}$.

In close analogy with vacuum quantum electrodynamics, zero-point fluctuations of fields in our one-dimensional system are controlled by the effective fine structure constant $\alpha = Z/R_Q$, where $R_Q = h/(2e)^2\approx 6.5~k\Omega$ is the resistance quantum for Cooper pairs~\cite{devoret1995quantum}. The superconducting state is favored for $\alpha \ll 1$, when the line mimics the usual weak-coupling electrodynamics of the free space, for which $\alpha = 1/137.0$. This  is not a coincidence: at a given frequency the Josephson relation links the fluctuation of phases across the junctions with the fluctuation of electric field between the chains, which in turn defines the strength of light-matter coupling. The model of a single chain coupled to a ground plane~(Fig. \ref{fig:Fig1}e) predicts a transition to the Mott insulator at $\alpha^{\textrm{Mott}}_c \approx 1/4$~\cite{korshunov1989effect, Choi1998Quantum}. The transition belongs to the celebrated Berezinski-Kosterlitz-Thouless (BKT) type and is driven by the competition between phase and Cooper pair number fluctuations.  
A more realistic model should include the oxide capacitance between neighboring islands and a random charge offset at every island. In this case the chain realizes a disordered Tomonaga-Luttinger liquid with the fine structure constant $\alpha$ replacing the inverse Luttinger interaction parameter~\cite{giamarchi2004quantum, bard2017superconductor, cedergren2017insulating}. In this case, theory predicts a compressible insulating state, termed ``Bose glass", at $\alpha^{\textrm{BG}}_c = 1/3$ ($Z= 2.2~k\Omega$)~\cite{PhysRevB.37.325}.
Although the critical point is formulated in terms of the wave impedance of the collective mode, little is known even in theory about what happens to this mode upon crossing the phase transition.
Traditional SI experiments focus on finding a universal scaling of (zero-frequency) DC resistance with temperature and other system parameters~\cite{lin2015superconductivity}. The quantum BKT transition resisted such an approach~\cite{chow1998length, bruder1999phase, haviland2000superconducting, ergul2013localizing}, possibly due to a combination of system's finite size, poor knowledge of actual temperature and microscopic parameters, or lack of equilibrium under a DC bias. In fact, disorder can make the resistance scaling on temperature non-universal~\cite{bard2017superconductor}. 
We avoid these issues by asking a conceptually different question: how the propagation of the phase mode, at frequencies higher than temperature, wavelengths much shorter than the system size, and excited with less than a single photon, is being inhibited in progressively higher impedance chains?

The minimal circuit model for our double-chain transmission line~(Fig. \ref{fig:Fig1}f) includes the junctions Josephson energy $E_J$, the charging energy $E_C = e^2/2C_J$, and the inter-chain coupling energy $E_0 = e^2/2C_0$. Here $C_J$ is the oxide capacitance between neighboring grains and $C_0$ is the capacitance between the two chains per unit cell. Introducing the unit cell size $a =600~\textrm{nm}$, the wave propagation parameters are now defined as $l\times a = 2(\hbar/2e)^2/E_J$ and $c\times a = C_0$. The junction plasma frequency $\omega_p\approx \sqrt{8 E_J E_C}/\hbar$ defines the ultra-violet cut-off in our system. The dimensionless parameter $(E_J/E_C)^{1/2}$ is proportional to the junction area. It defines the exponent of the quantum phase slip amplitude at the short length and time scales associated with $\omega_p$~\cite{bard2017superconductor}. 
With these parameters, the critical wave impedance matches that of a single chain~\cite{choi1998cotunneling}.

\begin{figure}
	\centering
	\includegraphics[width=\linewidth]{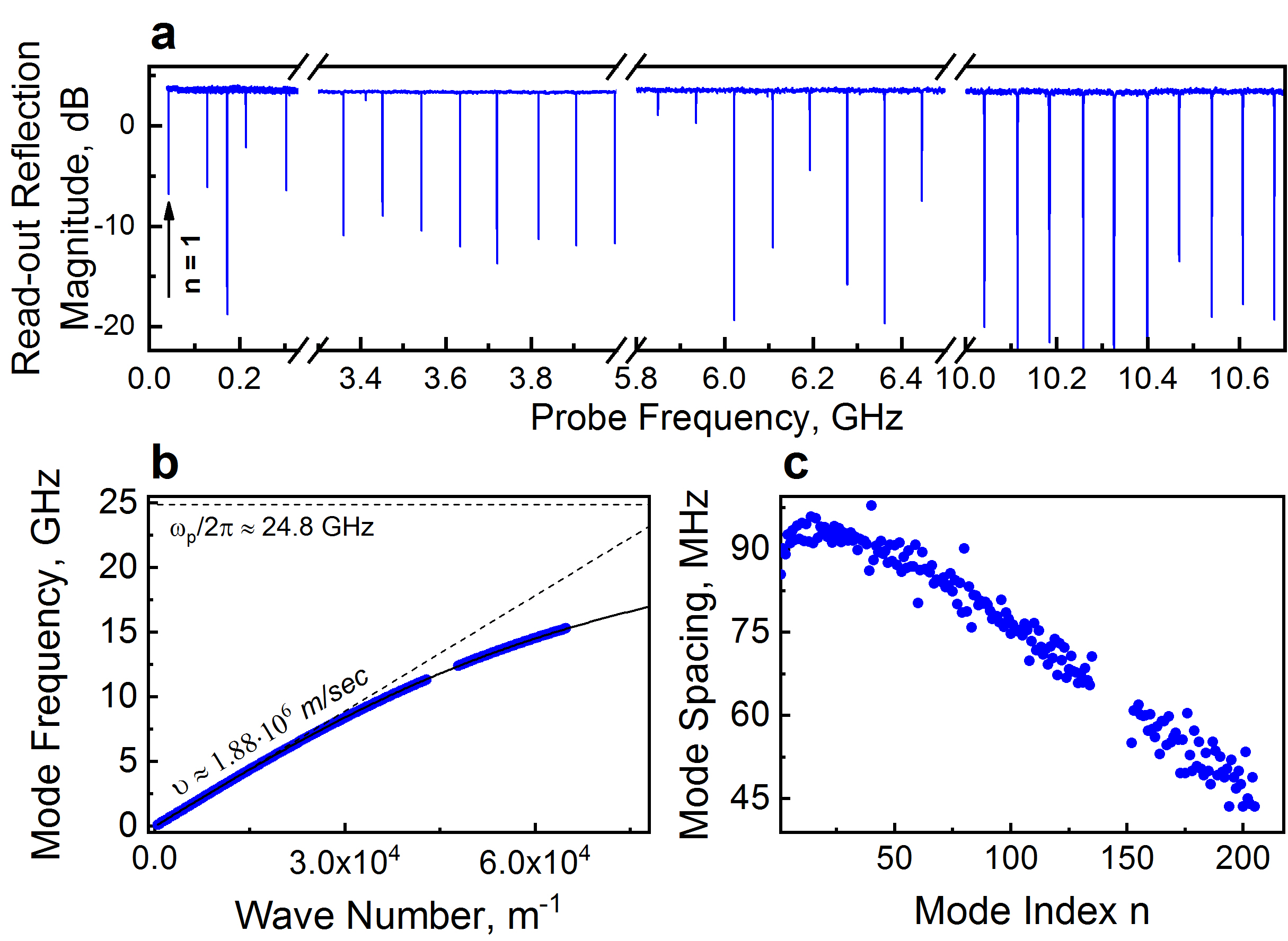}
	\caption{ (a) Reflection signal as a function of probe frequency. Discrete standing wave resonances are indexed one by one starting from the very first mode at about $40~\textrm{MHz}$. The third resonance is a spurious mode and is discarded. (b) Reconstructed dispersion relation (blue markers) and theoretical fit (solid line). (c) Mode spacing as a function of mode index. The sharp periodic outliers originate from the stitching error of the lithographer, otherwise invisible in device images.}

	\label{fig:Fig2}
\end{figure}

\begin{figure*}
	\centering
	\includegraphics[width=0.7\linewidth]{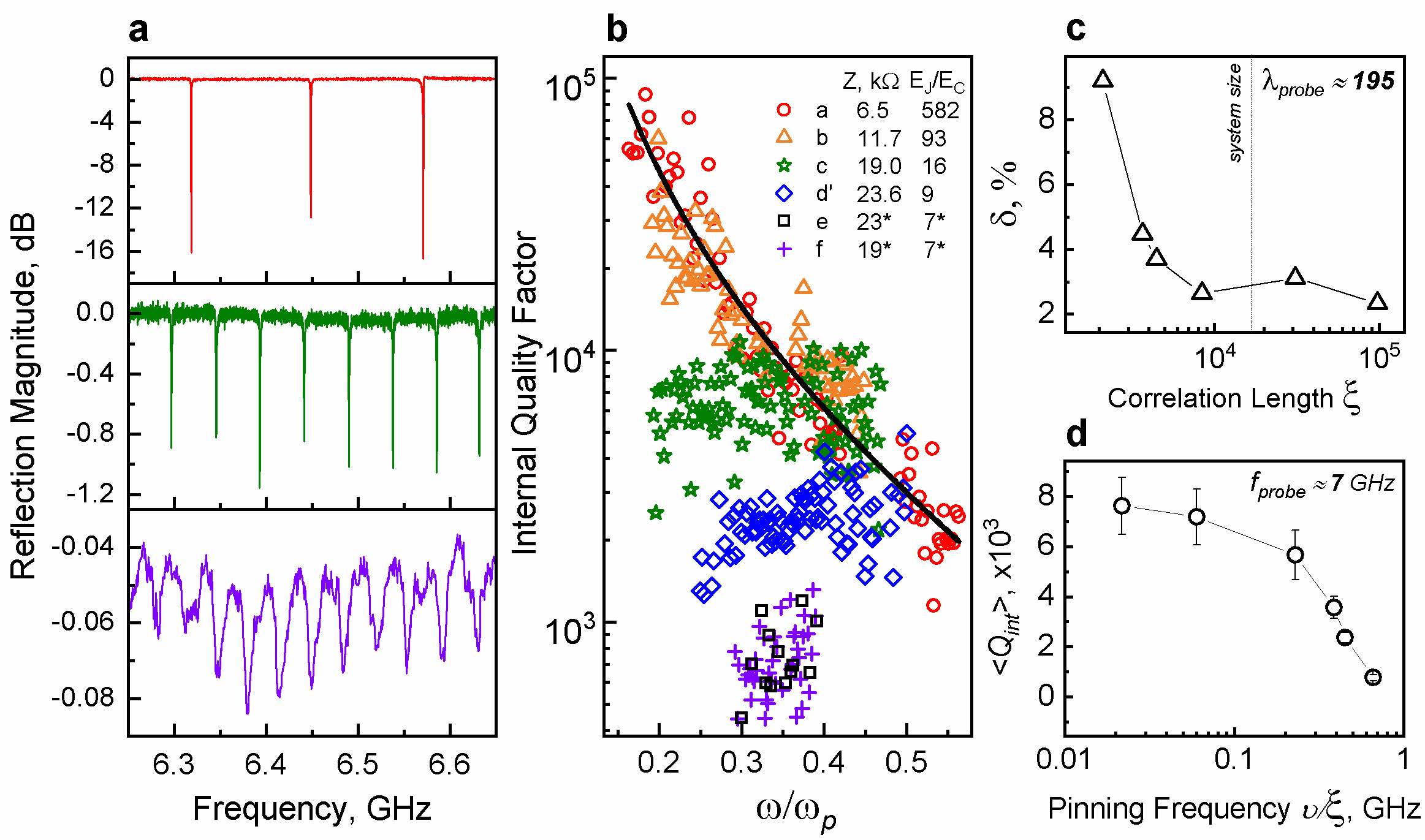}
	\caption{(a) Reflection magnitude (phase data not shown) vs. frequency for devices with reducing junction area. (b) Extracted $Q$-factor plotted as a function of the mode frequency normalized to the plasma frequency. (c) r.m.s mode spacing fluctuation at a wavelength $\lambda \approx 200$ unit cells, normalized by the average mode spacing, vs. theoretical correlation length $\xi$ of the 1D Bose glass. (d) Average quality factor at $\omega/\omega_p \approx 0.35$ (corresponds to $\approx 7$ GHz) vs. theoretical pinning frequency $v/\xi$.
	}
	
	\label{fig:Fig3}
\end{figure*}

An example of momentum-resolved spectroscopy is shown in~Fig. \ref{fig:Fig2}. 
The experiment is performed using a standard two-tone dispersive reflectometry, taking advantage of the weak Kerr non-linearity of a Josephson junction~\cite{weissl2015kerr} (Methods). Data reveals an ordered set of discrete resonances which we associate with the standing wave modes of the transmission line~(Fig. \ref{fig:Fig2}a). By indexing the individual resonances and plotting the frequency as a function of index $n = 1, 2,...$, we obtain the dispersion relation $\omega_n(k_n)$, where $k_n$ is the wavenumber defined as $k_{n+1} - k_{n} = \pi/L$, and $L = 10~\textrm{mm}$ is the length of the line~(Fig. \ref{fig:Fig2}b). The dispersion is in excellent agreement with a simple two parameter expression $\omega(k) = vk/\sqrt{1+(vk/\omega_p)^2}$, describing ultra-slow photons with a velocity $v =1.88\times 10^6~\textrm{m/s}$ and a band edge at the plasma frequency $\omega_p/2\pi = 24.8~\textrm{GHz}$. The $n=1$ mode is clearly visible at $40~\textrm{MHz}$, half the mode spacing, which correctly reflects the additional $\pi$ phase-shift due to the short-circuit boundary condition. This observation confirms that wave propagation occurs along the entire length of the system and the spectrum is gapless. Fluctuations of the mode spacing as a function of mode index are found to be within a few percent, independently of the wavenumber (Fig.~\ref{fig:Fig2}c), thus showing no signs of Anderson wave localization~\cite{basko2013disordered}.



The speed of light in the $10^6~\textrm{m/s}$ range requires an effective vacuum permeability of the order $10^4 \times \mu_0$. Such a large number can only come from the superconducting phase-stiffness of small-area Josephson junctions, which proves that we observed the phase mode. Combining the measured values of $\omega_p$, $v$, and the known dimensions of the chains, we reliably extract the wave impedance along with other device parameters (see Suppl. Mat.). For the device $b$ from Fig. 2, we get $Z = 11.7~k\Omega$. This is equivalent to a strikingly large value of the fine structure constant $\alpha = 1.8$, exceeding the theoretical insulator transition value by more than a factor of $5$. Yet, no signs of insulator are seen in the dispersion relation.


To reveal the insulating state we measured the linewidth of the collective mode resonances in the $4-12~\textrm{GHz}$ band for devices with progressively higher $Z$~(Fig. \ref{fig:Fig3}). The higher $Z$ was achieved by reducing the chain width, and hence the junctions area, while fixing all other dimensions~(Fig. \ref{fig:Fig1}b). We define the mode quality factor $Q$ as the ratio of mode frequency to its linewidth after subtracting the small contribution due to the radiation of photons into the measurement port (Methods). For devices with $Z \lesssim 12~k\Omega$ and $E_J/E_C \gtrsim 90$ the quality factor grows upon reducing the normalized frequency $\omega/\omega_p$ independently of other device parameters~(Fig. \ref{fig:Fig3}b - red and orange markers; see also Suppl. Mat.). Towards the lower end of the band, it reaches a relatively high value of $Q \approx 10^5$. This effect can be explained by the dielectric loss~\cite{martinis2005decoherence} in the tunnel barrier of the junctions~(Fig. \ref{fig:Fig3}b, solid line - see Suppl. Mat.). The key observation of this measurement is that a novel decoherence mechanism, with a reversed frequency-dependence of the $Q$-factor, takes over in weaker chains~(Fig. \ref{fig:Fig3}b - blue and violet markers). For some intermediate chain width (device c) the $Q$-factor becomes flat in frequency and it develops a clear tendency to drop towards lower frequencies in smaller width chains (devices d', e). In other words, the longer the wavelength of a phase mode excitation, the faster it decoheres. Such a behavior is highly unusual for materials-related loss, but it is consistent with the insulator state: by continuity, decoherence would suppress the DC transport ($\omega,k=0$). Although for technical reasons it was not possible to track the decrease of the $Q$-factor below $4~\textrm{GHz}$ (Methods), we checked that the low-temperature resistance of such devices exceeds $ 1~\textrm{M}\Omega$ and hence is indeed inconsistent with superconducting behavior (Suppl. Mat.).

The AC transport at $6-7~\textrm{GHz}$ becomes completely suppressed in chains with $E_J/E_C \lesssim 7$ and $Z \gtrsim 23~k\Omega$ ($\alpha \gtrsim 3$). There are two precursors to this: a sharp reduction in the $Q$-factor around a fixed frequency and a simultaneous increase of the spread in the modes spacing around a fixed wavenumber. To connect this data with existing theories of a 1D Bose glass insulator, we have calculated the correlation (pinning) lengths $\xi$ and the associated pinning frequency scales $v/\xi$ using the measured device parameters~\cite{fukuyama1978dynamics, giamarchi2004quantum, vogt2015one} (Suppl. Mat.). The r.m.s mode spacing fluctuation $\delta$ grows to about $10~\%$ upon reducing $\xi$ (Fig.~3c). The average $Q$-factor drops to about $5\times 10^{2}$ upon increasing the pinning frequency (Fig.~3d). 

\begin{figure}
	\centering
	\includegraphics[width=0.8\linewidth]{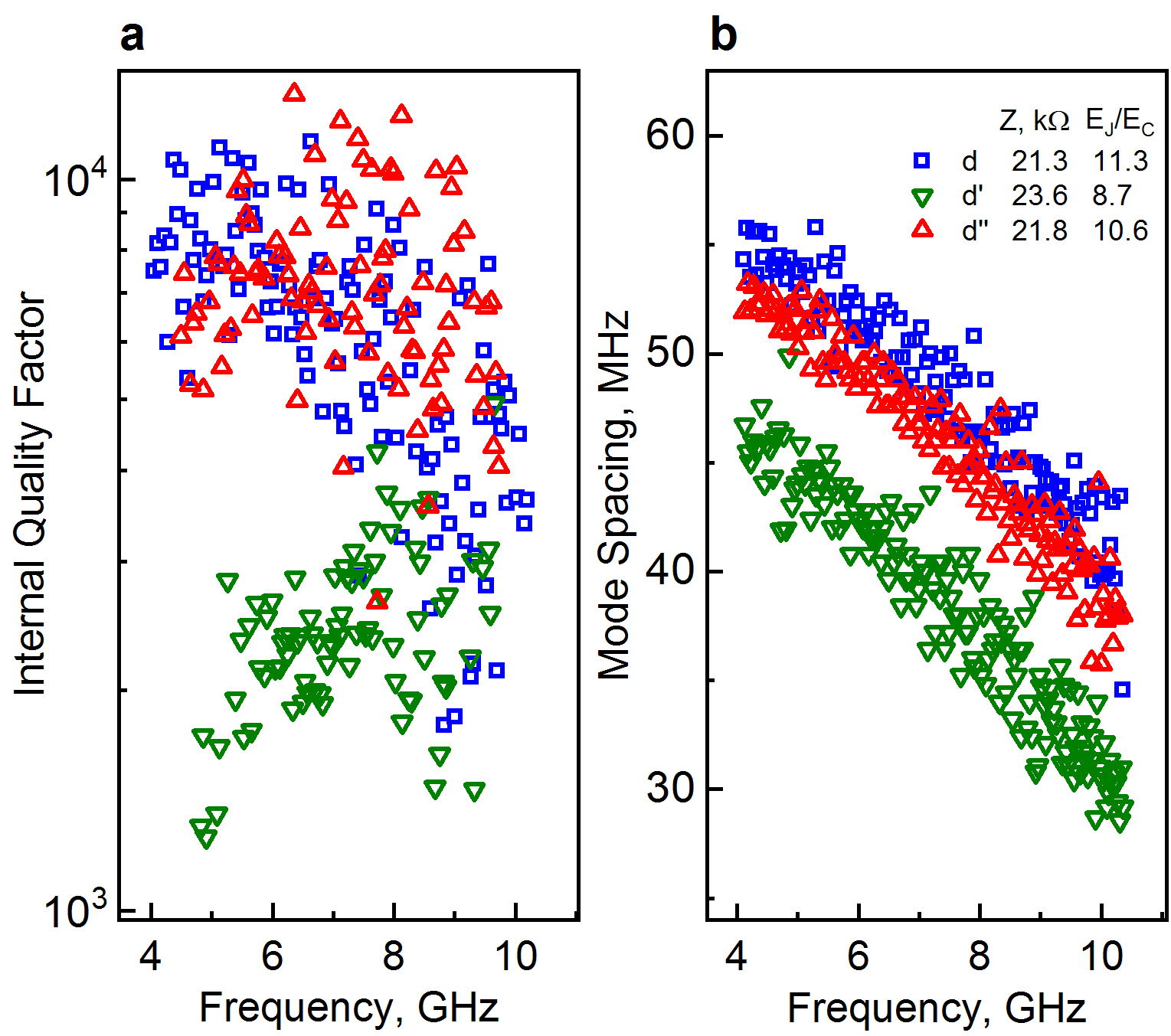}
	\caption{The $Q$-factor (a) and the mode spacings (b) as a function of frequency for the three subsequent incarnations of a single device: fresh after fabrication (blue), aged (green), and annealed (red). At $5~\textrm{GHz}$ the quality factor undergoes a remarkable swing by nearly an order of magnitude. }
	
	\label{fig:Fig4}
\end{figure}

In a key control experiment, we demonstrate a reversible transition between the two types of frequency dependence of the $Q$-factor in a single device (Fig.~4). A fresh device was fabricated with a low value of $E_J/E_C$ such that the measured $Q$-factor is still growing at low frequencies (Fig.~4a, device d). After aging for about 1000 hours at ambient conditions, the average $E_J$ reduced by about  $25\%$. This is confirmed by the mode spacing data clearly showing a reduction in both $v$ and $\omega_p$ (Fig.~4b, device d'). This small change in $E_J$ was enough to reverse the frequency dependence of the $Q$-factor. (Fig.~4a, device d'). Moreover, annealing the device at about $150\textrm{C}$ in ambient atmosphere recovered both the fresh value of $E_J$ together with the original frequency dependence of $Q$ (Fig.~4, device d''). Aging of devices with $E_J/E_C \gtrsim 70$ had no effect on $Q$. The high sensitivity of the $Q$-factor to $E_J/E_C$ (Fig.~3) and $E_J$ (Fig.~4) unambiguously links the observed decoherence to quantum phase slips~\cite{matveev2002persistent, rastelli2013quantum}. 
In the other control experiment we checked that reducing $Z$ by about $20\%$ without modifying $E_J/E_C$ -- achieved by shrinking the spacing between the chains from $10~\mu m$ to $2~\mu m$ -- showed no significant effect on the $Q$-factor (Fig.~3b, devices e vs. f).  Finally, we also checked that reducing the chain length by a factor of $10$ had no effect on the $Q$-factor, eliminating the finite-size effects (Suppl. Mat.).

Our observations can be broadly interpreted as the high-energy
view of a system near a quantum phase transition. All reported chains have parameters of an insulator in the conventional infra-red limit. Yet, they behave as superconductors when probed at wavelengths much shorter than the correlation (pinning) length $\xi$ and at frequencies much higher than the pinning frequency $v/\xi$ (Fig.~3c,d). Under such stimulus conditions, a disordered phase may be tricked to think that $\xi$ is effectively infinite. Hence it responds with a non-zero (but very low) phase-stiffness throughout the entire system length -- the necessary attribute of the ordered state. Importantly, the insulator leaves its footprint even in the high-energy limit. It consists of the small scatter (Fig.~3c) and broadening (Fig.~3d) of the collective mode resonances. These intrinsic interaction effects become visible when the values of $1/Q $ and $\delta$ outgrow the system-specific effects, such as dielectric loss and fabrication-induced disorder, respectively. We demonstrated this by either reducing $\xi$ with weaker chains or by raising the excitation wavelength. 

Decoherence of the phase mode in a 1D Bose glass has been explored theoretically in response to our experiment~\cite{wu2018theory, bard2018decay, houzet2019microwave}. One mechanism is that a single probe photon decays into lower-frequency photons~\cite{wu2018theory}. The other one consists of inelastic collisions of the probe photon with low-frequency thermal photons~\cite{bard2018decay}. The third scenario describes the pinning of a charge density wave, which is a dual representation of the phase mode~\cite{houzet2019microwave}. It causes an inhomogeneous broadening of the standing wave resonances after ensemble-averaging of their frequencies over disorder. This can indeed occur in a single device due to fluctuating in time offset charges. All models yield a growing value of $1/Q$ at lower frequencies, qualitatively similar to our observations. The mode frequency scatter $\delta$ (Fig.~3d) remains unexplained.
These initial theories illustrate the advantage of the energy knob: by increasing the probe frequency, we effectively reduce the interactions in the renormalization group sense, and hence may keep the interaction-induced measurable quantities $\delta$ and $1/Q$ small, such that they can be calculated perturbatively. This will help bridging theory and experiment of the SI transition in a model BKT system. Our approach can be further extended to more complex SI transitions in thin superconducting films~\cite{crane2007survival} and nanowires~\cite{ku2010superconducting}, where DC measurements left many fundamental questions unresolved~\cite{vinokur2008superinsulator, arutyunov2008superconductivity, sacepe2011localization, feigel2018microwave}. 

Irrespective of the interpretation, a 1D electromagnetic mode with $\alpha \sim 1$ is a unique resource for quantum science and technology. Any atom with a transition dipole in the microwave band will sense quantum fluctuations of the electric field in the gap between the two chains (see Fig.~1a). 
For a dipole size matching the gap ($\sim 1~\mu\textrm{m}$), the condition $\alpha \sim 1$ means the spontaneous emission lifetime is comparable to the atomic transition cycle, a non-perturbative situation where the traditional QED fails. In principle, the required size dipole can be provided even by a single Rydberg atom slowed down near our on-chip superconducting waveguide~\cite{haroche2006exploring}. 
More generally, the availability of 1D  microwave photons with impedance over $23~\textrm{k}\Omega$, demonstrated here for the first time, can transform many hybrid quantum platforms: from trapped polar molecules~\cite{andre2006coherent} and electrons on helium surface~\cite{schuster2010proposal} to fabricated quantum dots~\cite{stockklauser2017strong} and nanomechanical systems~\cite{arrangoiz2018coupling}.
\newpage
\noindent\textbf{METHODS}\\
\\
More details, including a table of all measured devices and their extracted parameters, are in the Supplementary Material.\\
\\
\textbf{Device fabrication}\\
The chains were fabricated using the standard Dolan bridge technique involving a MMA/PMMA bi-layer resist patterned by electron beam lithography with subsequent double-angle deposition of aluminium with an intermediate oxidation step. The substrate is a high-resistivity silicon wafer. Due to the large number of junctions in the chain, patterning was done by stitching multiple fields of view with a size of $100~\mu m$. The stitching error is invisible in device images. Curiously, it can be clearly seen in Fig.~\ref{fig:Fig2}c as sharp periodic shifts in the mode spacing data. We use this information to confirm the conversion between the standing wave index and the wavelength.\\
\textbf{Wireless RF-spectroscopy setup}\\
The chip hosting the chains is mounted at the center of a copper waveguide (Fig.~\ref{fig:Fig1}c). In order to launch microwaves we have designed a coaxial-to-waveguide transition launcher with a good matching in the range $7-12~\textrm{GHz}$. The antenna attached to the chain is smaller than the free space wavelength at these frequencies. Therefore, the combination of the chip antenna, the waveguide box, and the launcher can be viewed as a semi-transparent``mirror" with a frequency-dependent finesse. Note that the two chains of the transmission line are spaced by only a few micrometers, whereas the distance between a chain and a wall of the copper box is at least $5~\textrm{mm}$, comparable to the full length of the chain. Such a setup minimizes typical parasitic capacitances due to the measurement circuitry seen by the junctions. The setup was cooled down in a dilution refrigerator with a base temperature around $10~\textrm{mK}$, which corresponds to a frequency of $200~\textrm{MHz}$ and thermal length $\approx L/2$. However, in numerous superconducting qubit experiments measured in a similar setup, the circuit temperature was often found to be closer to a $50~\textrm{mK}$, which would correspond to a proportionally reduced thermal length.\\
\textbf{One-tone spectroscopy}\\
We used a Rohde \& Schwarz ZNB network analyzer to measure the frequency-dependent reflection amplitude and phase in a single-port reflection experiment. To fit the data in the vicinity of each resonance, we use the commonly known expression:
\begin{equation}
S_{11}(\omega) = \frac{2i(\omega-\omega_0)/\omega_0- Q^{-1}_{\textrm{ext}}+Q^{-1}_{\textrm{int}}}{2i(\omega-\omega_0)/\omega_0 + Q^{-1}_{\textrm{ext}}+Q^{-1}_{\textrm{int}}},
	\end{equation}
where $\omega_0$ is the resonance frequency, $Q_{int}\equiv Q$ is the internal quality factor, plotted in Figs.~{3-4}, and $Q_{\textrm{ext}}$ is the external quality factor, which in general is a complex number. It's real part can be viewed as a measure of opacity of the mirror at the antenna end of the chain. We found that at frequencies below $7~\textrm{GHz}$, the opacity grows upon reducing the frequency which is consistent with the propagation cut-off of our copper waveguide. The opacity also has a tendency to increase as frequency grows above $10~\textrm{GHz}$ which may be related to a partial Anderson localization of microwaves in the chain and their decoupling from the antenna end. As a result, one-tone spectroscopy becomes inefficient far outside the $7-12~\textrm{GHz}$ pass-band of our coaxial-to-waveguide launching system. For this reason, the frequency range in Fig.~\ref{fig:Fig3} is limited. Finally, we note that our reflection data fits exceptionally well to the above expression for $S_{11}(\omega)$, as described in the Supplementary material, which allows very accurate extraction of both $\omega_0$ and $Q$.\\
\textbf{Two-tone spectroscopy}\\
We use a weak cross-Kerr interaction between the photons in different modes in order to perform broadband spectroscopy shown in Fig.~\ref{fig:Fig2}. First, the readout mode is selected in the pass-band. Reflection amplitude and phase at a properly chosen frequency near the resonance are measured as a function of the frequency of the second tone, which is scanned to look for other modes. The cross-Kerr effect results in the shifting of the frequency of the readout mode due to the photon occupation of every other mode in the system. Since expressions for frequency shift per photon are readily calculable, we use this information to approximately calibrate the measurement power down to a single-photon level. In chains with $E_J/E_C \approx 10$, the spectrum below a frequency of about $1~\textrm{GHz}$ becomes difficult to interpret. Nevertheless, there are resonances down to a frequency of about $100~\textrm{MHz}$.\\
\textbf{Extracting $Z$ from dispersion relation}\\
We tried two methods for extracting $Z$. First is based on our knowledge of junction areas and the fact that the oxide capacitance has a rather device-independent value of $45~\textrm{fF}/\mu m^2$. The second method is based on the known formulas for the capacitance of two infinitely long coplanar strips. Both methods yield consistent results within $20\%$. We used the more conservative result (smaller values of $Z$ and higher values of $E_J/E_C$) in the main text of the manuscript.

\bibliography{SuperconductingCircuits}
\bibliographystyle{naturemag}
\end{document}


\title{Supplementary Material for\\ ``Quantum electrodynamics of a superconductor-insulator phase transition''}
\maketitle
\captionsetup[figure]{font=small,labelfont={bf},labelsep=period,labelformat={default},name={Fig. S}}

\section{Samples}

\begin{table}[htbp]
  \centering
  \caption{Josephson transmission line parameters}
      \begin{tabular}{c|c|c|c|c|c|c|c|c|c}
          &       &       &       & \multicolumn{2}{p{7.37em}|}{Method 1 (Junction area)} & \multicolumn{2}{p{7.785em}|}{Method 2 (Stripes capacitance)}        &       &\\
    \midrule
    Device & \multicolumn{1}{p{3.08em}|}{$v, 10^6$ m/sec} & \multicolumn{1}{p{3.185em}|}{$\omega_p/2\pi$, GHz} & \multicolumn{1}{c|}{$\Lambda$} &\multicolumn{1}{p{3.29em}|}{$Z$, kOhm} & \multicolumn{1}{p{3.08em}|}{$E_J/E_C$} & \multicolumn{1}{p{3.475em}|}{$Z$, kOhm} & \multicolumn{1}{p{3.21em}|}{$E_J/E_C$} & \multicolumn{1}{p{3.08em}|}{$\xi$, unit cells} & \multicolumn{1}{p{3.21em}}{$v/\xi$, MHz} \\

     \midrule
	$g$     & 22.6 & 26.7  & 4.7   & 0.7   & 712   & 0.7   & 745 & $--$ & $-$ \\
    \midrule
    $h$     & 8.20   & 21.5  & 8.4   & 2.3   & 211 & 2.2   & 237& $--$ & $-$  \\
    \midrule
    $i$     & 2.11  & 20.9  & 37.9  & 7.0     & 484 & 7.6   & 411 & $\gg 10^6$ & $\ll 1$ \\
    \midrule
    $a$     & 2.76  & 27.0    & 38.4  & 7.4   & 440 & 6.5   & 582 & $\gg 10^6$ & $\ll 1$ \\
    \midrule
    $b$     & 1.88  & 24.8  & 28.4  & 12.6  & 80  & 11.7  & 93  & $\gg 10^6$ & $\ll 1$ \\
    \midrule
    $j$     & 1.68  & 22.3  & 28.2  & 13.8  & 65  & 13.1  & 72  & $\gg 10^6$ & $\ll 1$ \\
     \midrule
    $k$    & 1.28  & 23.8  & 20.2  & 17.0    & 20.2  & 20.3  & 13.7  & 97815 & 22 \\
    \midrule
    $c$     & 1.12  & 20.8  & 20.2  & 19.0    & 15.8  & 23.1  & 10.2  & 31023 & 60 \\
    \midrule
     $d$     & 1.16  & 22.3  & 19.4  & 21.3  & 11.3  & 23.8  & 8.7   & 8399  & 230 \\
    \midrule
    $l$      & 1.04  & 20    & 19.4  & 23.1  & 9.3   & 26.6  & 6.7   & 4461  & 387 \\
    \midrule
     $d'$     & 0.99  & 19.3  & 19.3  & 23.6  & 8.7   & 27.8  & 5.9   & 3655  & 452 \\ 
    \midrule
    $d''$     & 1.12  & 21.6  & 19.4  & 21.8  & 10.6  & 24.7  & 8     & 6837  & 272 \\
    \midrule
    $e$     & 0.98*  & 20.8*  & 17.7*  & 23*  & 7*   & 29*  & 5* & 2348  & 699 \\
    \midrule
    $f$     & 0.83*  & 20.8*  & 15*    & 19*  & 7*   & 22*  & 5*& 2105  & 659  \\
    
    \end{tabular}%
  \label{tab:addlabel}%
\end{table}%

All lines are 10 mm long with the unit cell size $a=$ 0.6 $\mu$m. The exceptions are devices $g$ and $h$, where $a$ is 40.2 $\mu$m and 10.2 $\mu$m respectively. The devices $k, c, d, l, d'$ and $f$ are used in the Fig.3 c,d. The correlation (pinning) length $\xi$ is found using the known device parameters as $\xi \approx (4E_0/W)^{2/(3-R_Q/Z)}$, where $W = 16/\sqrt{\pi}(8E_J^3E_C)^{1/4}\exp{-\sqrt{8E_J/E_C}}$.

\section{Two-tone spectroscopy}
The direct reflection measurement is limited in our RF setup to approximately 4 - 12 GHz bandwidth. The lower bound is defined by the strong attenuation of the 3D copper waveguide having its cut-off near 7 GHz. The upper bound is given by the pass band edge of the isolators and that happen to be at 12 GHz. In order to reveal as many collective modes as possible momentum resolved two-tone spectroscopy was used. This technique exploits a weak non-linearity of the Josephson junctions array. The quartic term in the junction potential gives rise to the so called cross-Kerr interaction between different collective modes. The interaction appears as a dispersive shift $\delta f_i=\chi_{ij}\cdot n_j$  of mode $i$ linear in photon population $n_j$ of any other mode $j$, with the linear coefficient given by $$\chi_{ij}\sim\frac{1}{N}\frac{f_i f_j}{E_J}$$
In a typical experiment we choose a mode in the 4 – 12 GHz range and fix the first read-out tone frequency at its left shoulder (Fig. S 1a). Another generator is used to apply the second RF tone which we call the probe tone since it probes frequency ranges inaccessible through direct reflection measurement. While the reflection of the read-out tone is being measured the probe tone frequency is swept. Every time the probe tone populates any of the collective modes the read-out reflection magnitude drops due to the downward shift of the read-out mode (Fig. S 1b). We can now plot the measured read-out reflection magnitude as a function of the probe frequency to obtain the spectroscopy shown in Fig. 2a. In all two-tone experiments the probe and read-out tone powers were always chosen to get the smallest possible shift usually less than or comparable to the read-out mode linewidth.

\begin{figure}[h]
\centering
\includegraphics[width=1\textwidth]{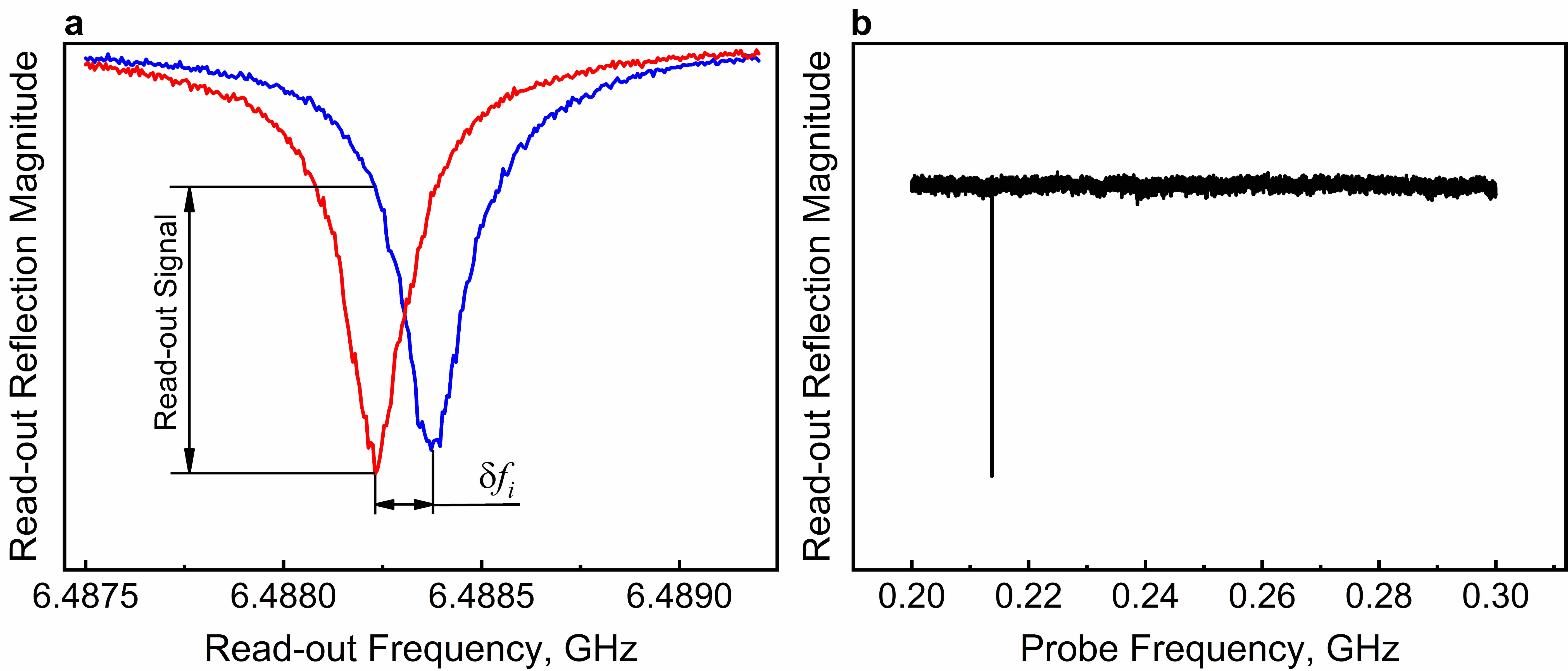}
\caption{(a) The read-out mode while the probe tone is in (red) and out off resonance (blue) with the collective mode at 0.215 GHz. (b) The read-out tone reflection magnitude measured at the left shoulder of read-out mode line as a function of the probe tone frequency. }
\end{figure}

The two-tone experiment has an advantage of allowing RF power calibration. The measured read-out mode shift can be compared to the known cross-Kerr coefficient to obtain an approximate correspondence between the photon number reaching the Josephson transmission line and the power we send into our RF input line. For example, the shift at Fig. S 1a corresponds to an order of 10 photons populating the mode at 0.215 GHz.  We use this information in the other experiments.

\section{Extraction of the transmission line impedance}
With the help of two-tone technique collective modes of the transmission line are usually found starting with the very first one. The modes now can be counted with a wavenumber assigned to each of them $k_n = 2\pi/L+\pi n/L$. Fig. 2b shows that the dispersion relation for plasmons in the telegraph like transmission line (Fig.1f) perfectly describes the measured collective mode frequencies. The least-squares residuals are usually randomly distributed around the fitting curve. Their small values ($<$ 10 MHz) and the large number of measured modes ($>$ 100) allow the precise extraction of two propagation constants, namely plasmon velocity $v$ and cut-off frequency $\omega_p$ (Table 1). With known junction length $a$ these constants can be related to the Josephson and charging energies $$v = 2a\sqrt{E_J E_0}/\hbar$$$$\hbar\omega_p = \sqrt{8E_J E_C} - E_C$$
The anharmonicity of the junction was taken into account in the last equation which is in particularly important for weaker chains with small $E_J/E_C$ ratio where the non-linearity is strong.

Table 1 also contains the values for the screening length of the charge-charge interaction inside our chains $\Lambda=\sqrt{E_0/E_C}$. The numbers can be readily obtained from the ratio of $v$ and  $\omega_p$ while neglecting the anharmonicity term which correction to $\Lambda$ is small. 

In order to find the line impedance $Z = 2/\pi R_Q\sqrt{E_0/E_J}$ as well as the Josephson and charging energies based on the measured $v$ and $\omega_p$ we used two separate methods. The first one exploits the fact that the Al oxide growth is typically self-terminating. It implies that the Josephson junction capacitance is fully defined by the junction area. We use SEM to measure the junction areas along the chain. The $C_J$ is found then as an average junction area multiplied by an empirical constant  $45 fF/\mu m^2$. The second method of finding the wave impedance takes the known geometrical dimensions of the chains and uses the well-known result for the capacitance of the coplanar stripes in order to calculate $C_0$. These methods give the impedance values which are different from each other by no more than 20\% (Table 1). This number can be considered as our accuracy for an absolute impedance measurement. The relative changes of $Z$ and $E_J/E_C$  can be tracked much more precisely for a single device when neither $C_J$ nor $C_0$ changes.

The most of the devices we measured revealed more than 100 modes in a wide frequency range. This allowed us to reconstruct the dispersion quite well and extract the $v$ and $\omega_p$ with a 1\% precision. However due to the strong decoherence discussed in the main text, the weakest chains did not show enough number of modes to accurately fit the theory. For such devices the parameters were estimated based on the mode spacing, known junction dimensions, and the expected plasma frequency obtained from measurements of multiple closely related devices. The asterisk in the Table 1 signifies the parameters obtained not from the dispersion fitting.

\section{$Q$ factor measurements}
In order to find the plasmon waves decoherence we performed the quality factor measurements for all the collective modes accessible from one-tone spectroscopy. The magnitude and phase of the reflection coefficient near the resonances were accurately measured at single photon power. Each resonance was then fitted to a model of a dissipative LC-oscillator with the expression (1) in Methods (Fig. S 2). 
As a result three parameters were extracted: the resonant frequency $\omega_0$, the internal $Q_{int}$ and external $Q_{ext}$ quality factors. $Q_{int}$ is the measure of decoherence inside the transmission line at $\omega_0$. $Q_{ext}$ can be defined as an inverse decay rate of the mode photons into the input-output port and it tells us how strongly each particular mode is coupled to the measurement apparatus.

\begin{figure}[!htbp]
\centering
\includegraphics[width=0.7\textwidth]{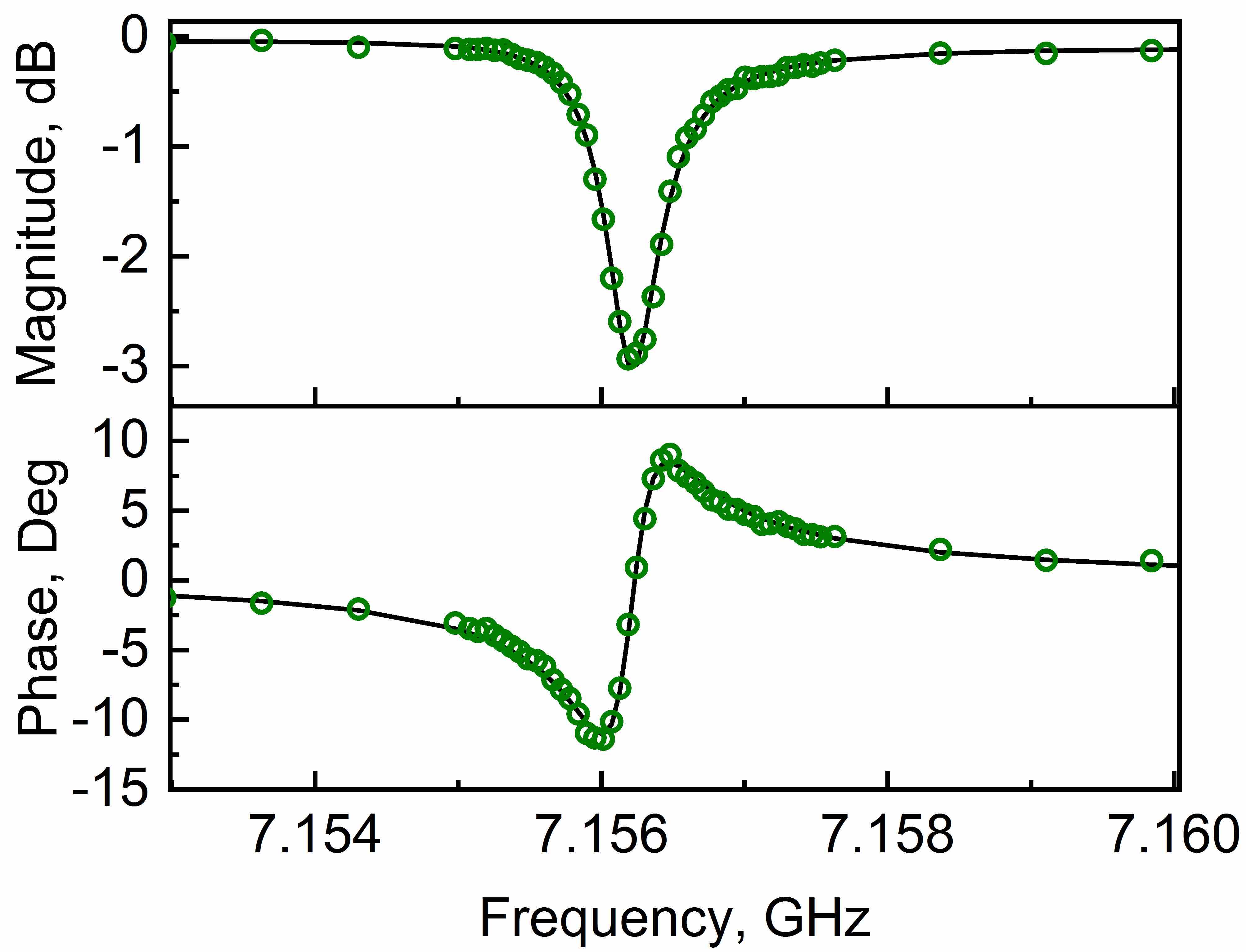}
\caption{A typical example of the reflection coefficient magnitude and phase for one of the collective modes. The data (green circles) is fitted to a model of dissipative LC-oscillator (black line).}
\end{figure}

\begin{figure}[!htbp]
\centering
\includegraphics[width=0.7\textwidth]{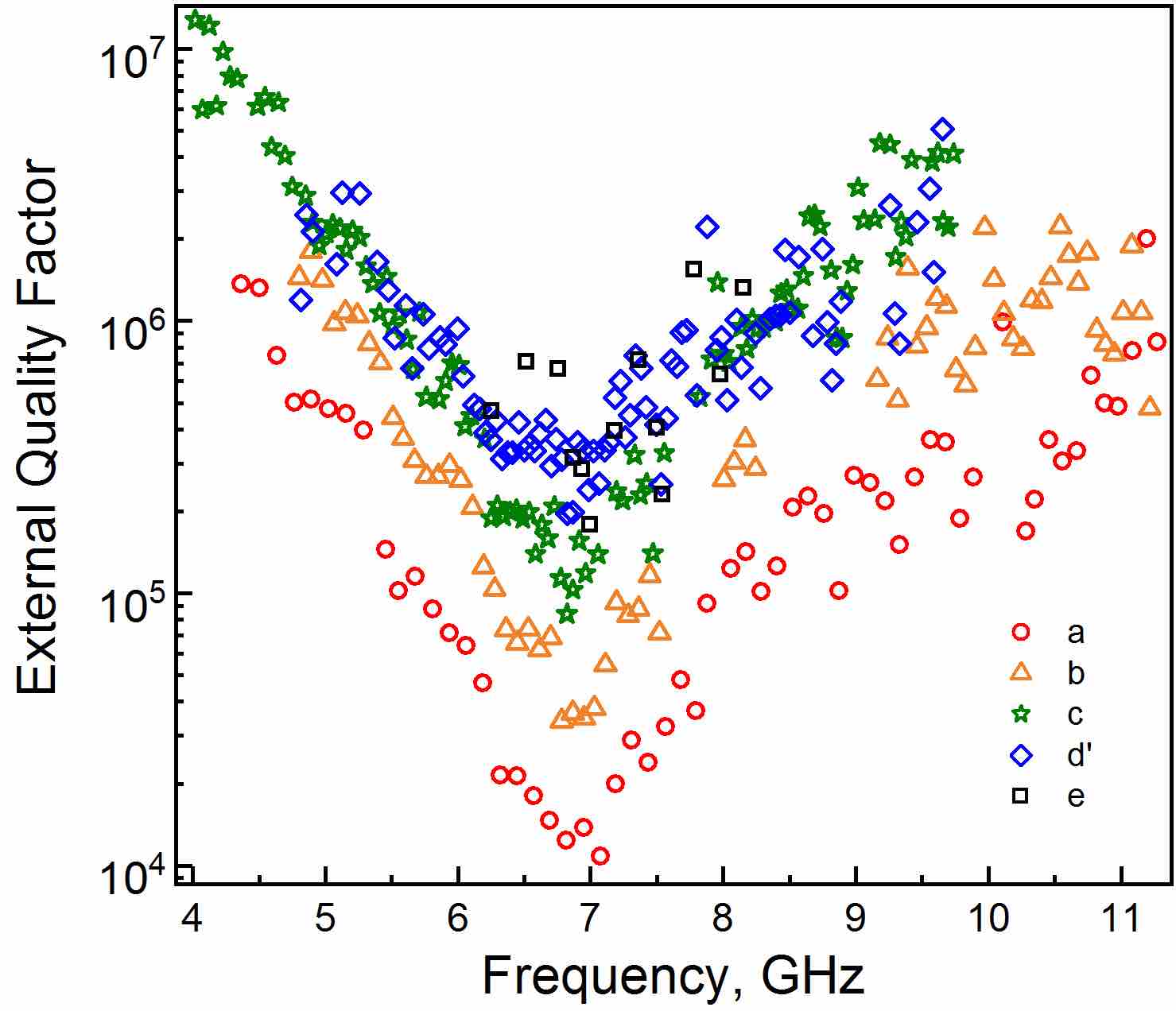}
\caption{External quality factor as a function of frequency. The common increasing trend below 7 GHz is the effect of 3D waveguide attenuation. The reduction of chain width (a to e) affects $Q_{ext}$ of all measured modes but more significantly the higher frequency ones.}
\end{figure}

Typically in our experiment the external quality factor is much higher than the internal one. This makes its impact to the total quality factor $1/Q=1/Q_{int}+1/Q_{ext}$ quite small. We therefore refer to $Q_{int}$ as just the quality factor $Q$ in the main text. The mode linewidth becomes an intuitive indicator of the decoherence strength. Very good agreement between our data and fit allows us to extract the quality factors within 5-10\% precision. The much larger spread in $Q$ seen at Fig. 3 and 4 is a property of the system and it is not related to fitting or measurement procedures.

The RF setup bandwidth is not the only limiting factor for our $Q$ measurements. The number of modes whose quality factors are shown in Fig.3 were also limited for the following two reasons. First of all the internal quality factor drops at higher frequencies for wider chains and at lower frequencies for thinner chains as was explained in the main text. This broadens the resonances making their measurement harder. The second and the most important reason is the rapid growth of external quality factor for weaker and weaker chains (Fig. S 3). This surprising finding limited the amount of modes accessible for direct reflection measurements. Strong suppression of modes coupling seen in the narrow chains seems like a proliferation of some localization effect. Thought more studies are required the observed behavior possibly indicates an important role of disorder in our system. It places the Josephson transmission line metamaterial on the intriguing intersection of superconductivity, strong interactions and localization.

\section{Decoherence in the low-impedance lines}
To test if the decoherence behavior observed for the lines with $Z\approx$ 7 kOhm persists at $\alpha < \alpha_c$ we studied several low-impedance lines. The lines were fabricated by increasing the length of the aluminium islands while keeping constant the deposition angles and the total device length. As a result the inductance per unit length goes down making the impedance lower. We found no noticeable change in the decoherence behavior of these low-impedance transmission lines (devices g and h at Fig. S 4).

\begin{figure}[!htbp]
\centering
\includegraphics[width=0.55\textwidth]{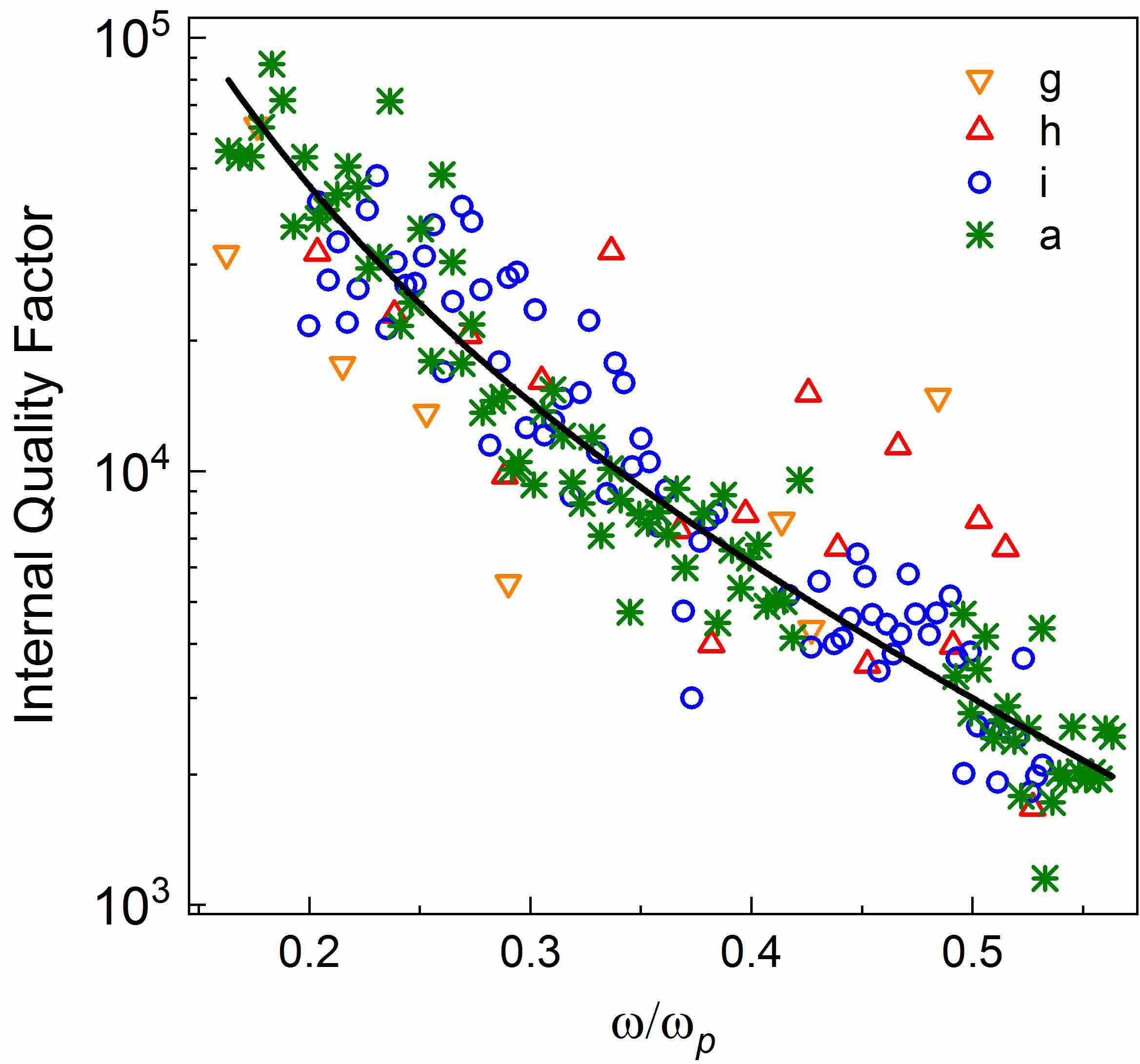}
\caption{Internal quality factor as a function of the mode frequency normalized to the plasma frequency measured for several low-impedance Josephson transmission lines. Solid black line is the prediction for the internal quality factor limited by a dielectric loss in junctions oxide.}
\end{figure}

The suppression of the quality factor towards the plasma frequency can be modeled by a dielectric loss in the junctions oxide (solid black line at Fig.S4). The model assumes a small imaginary part of the junction capacitance $C_J$ and it predicts the universal dependence of $Q_{int}$ on the normalized frequency $$Q_{int}\approx\frac{2}{\tan\delta}\left(\frac{\omega_p}{\omega}\right)^{2}\left(1-\left(\frac{\omega}{\omega_p}\right)^2\right)$$, where the loss tangent is either a constant $\tan\delta = \frac{Im  C_J}{Re  C_J}$, or a function slowly growing with frequency $\tan\delta \sim \omega^{0.7}$. By fitting our data to the above expression we obtained $\tan \delta \sim 10^{-3}$ which is consistent with the loss tangent of AlOx.

\section{Length dependence in RF and DC experiments}
When $N\to\infty$ and $T\to 0$, theory offers a straightforward separation between a superconductor and an insulator as the states with zero and infinite resistance. If both $N$ and $T$ are finite, which is always true, the state identification becomes a challenge. 
The DC resistance $R_0$ is now finite in both cases and it depends on $N$ and $T$. Its dependence on $N$ is predicted to be strongly non-monotonic as far as $N\ll N_{th}$, where $N_{th}$ is the thermal length. 
When $N\gg N_{th}$, the dependence becomes ohmic with the resistivity controlled by a temperature in a non-monotonic way as well. The picture is further complicated by the fact that the actual device temperature and hence $N_{th}$ is actually unknown because a superconducting circuit is usually not in thermal equilibrium with the bath. 

\begin{figure}[!htbp]
\centering
\includegraphics[width=0.55\textwidth]{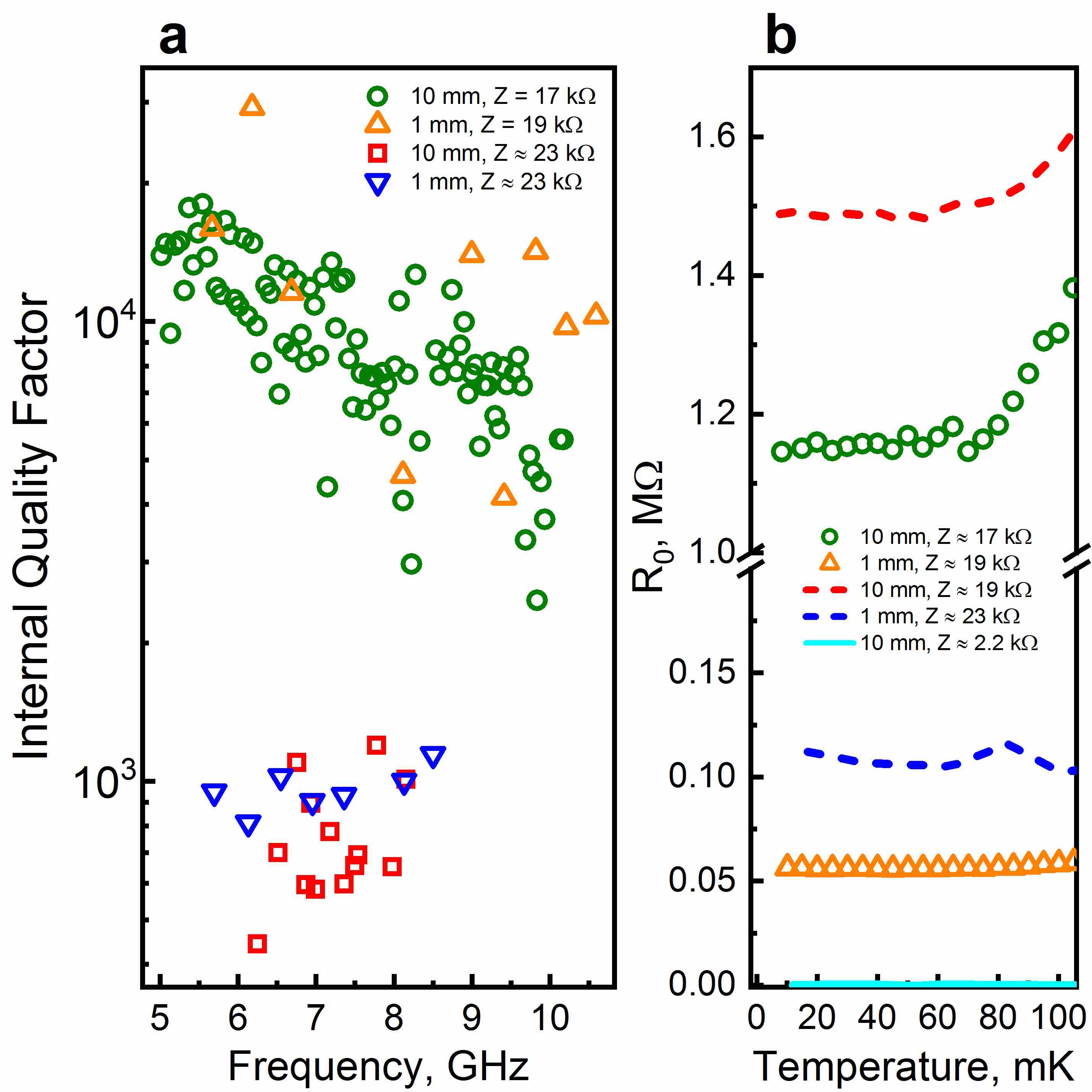}
\caption{(a) Internal quality factor as a function of mode frequency normalized to the plasma frequency measured for two pairs of nominally identical transmission lines but with the 10x difference in the length. (b) Zero-biased DC resistance as a function of temperature for several transmission lines of varying length and impedance.}
\end{figure}

To see the role of the $N$ in our experiment and to check the system's behavior in $\omega\to 0$ limit, we probed the RF and DC responses of several Josephson junction transmission lines of varying length.
We found that the internal quality factors show no significant change between nominally identical devices of varying length (Fig. S5 a). Thus the quality factors in the weak chains are still suppressed and tend to lower values as the frequency is decreased. The absence of length dependence is not a surprise here since the wavelength $\lambda$, being much smaller than $N$ and $N_{th}$, is now an important length scale. 

The situation is less transparent when we look at the DC resistances of the same devices (Fig.S5 b). The measurements were done at currents 1-10 pA using a standard lock-in technique. All measured devices showed the flattening of their DC resistance at the lowest temperatures, which is typically observed in Josephson junction arrays. The resistance value is found to be depended on the transmission line impedance. The resistance is around zero when $Z\approx Z_c$ (cyan line at Fig.S5 b) and grows, reaching over 1 M$\Omega$ values, as $Z$ is tuned deeper into insulating state. 
Comparing the devices of varying length we found that the resistance growths faster than $N$. This suggests that the arrays are indeed in an insulating state as the theory predicts, though it is hard to be absolutely certain when the actual thermal length is unknown, and more thorough DC studies need to be done.

By comparing RF and DC experiments we can come to the following conclusions: 
1) Over 1 M$\Omega$ DC resistances measured in 10 mm long transmission lines are clearly inconsistent with wave propagation. This follows from a simple fact that a transmission line with a M$\Omega$ resistance would have $Q\ll 1$. This confirms our conclusion that the waves survive even when the DC transport is suppressed.
2) The absolute value of DC resistance tells nothing about the system state. The $Q_{int}$ being length independent offers a direct characterization of how strong an insulating or superconducting state is at a particular set of parameters.
3) The temperature range in a DC experiment with Al/AlOx/Al arrays where existing theory can be applied is small since it is limited by a thermalization problem at one side and quasiparticle generation at the other side. In contrast, the frequency range available for the quality factor measurements can be much wider (up to three orders of magnitude in $\omega$). It gives an advantage to the finite frequency studies of the SI physics.

\section{Aging}
The fine tuning of $E_J/E_C$ ratio used to reveal the transition at Fig.4 was achieved with the help of aging. The effect of aging relies on increasing of the junction normal resistance with time spent by the junction in the air. The increasing of the normal resistance results in growth of the Josephson inductance, which can be observed in our experiment by measuring the cut-off frequency of the plasmons dispersion. The process of aging is self-terminating so the plasma frequency saturates in several weeks following an almost exponential decay in the first few hours.


The Josephson inductance during the aging changes uniformly along the whole chain which manifests itself as the uniform shift of modes spacings so
every single mode experiences a proportional to its mode index downward shift in frequency (Fig. 4b). The resilience of the mode spacing fluctuations to aging observed in many our devices allowed us to relate them to a disorder in the chain parameters and to use their values to estimate the standard deviation of the junction areas. It typical values are in the 5-10\% range. By measuring multiple devices with the values of $E_J/E_C>70$ we can confirm that the aging does not affect the quality factors of their modes.

The effect of aging is not permanent and can be reversed. We found that two hours baking of the Josephson transmission line on a hot plate at ambient conditions can reverse almost 1000 hours of aging. This resulted in resurrection of almost the original chain parameters and modes quality factors, as shown in the experiment demonstrated in Fig. 4. It worth noting that the external quality factors were not affected in this experiment.

\section{The spectroscopy in the weakest chains}
In weaker chains the two-tone spectroscopy stops working at low frequencies. The first reason for that is the read-out mode becoming broad and shallow reducing the signal to noise ratio. However the most important reason is the population of low frequency modes becomes harder and harder in weaker chains (Fig.S 5a). As a result low frequency modes eventually disappear from the two-tone spectroscopy. This behavior is in agreement with the internal quality factor reduction at low frequencies observed at higher frequencies with the help of one-tone reflectometry. The growth of external quality factor can also play here a role. To find the mode indices in the situation when not all modes can be found we used an extrapolation from the higher modes.

The second noticeable effect taking the place in the weakest chains is related to their modes frequencies. The high distortion in the collective mode positions was observed in many weak chains. The observation appears as the large fluctuations in the mode spacings as a function of frequency (Fig.S 5b c). The spacings also sometimes take unexpected values which are inconsistent with the planned device parameters. The described effect seems to be very sensitive to small changes in the fabrication process and may or may not appear even in nominally identical devices. To characterize the spread in the modes positions we use the quantity $\delta$ defined as $$\delta=\frac{1}{\overline{\Delta f}} \sqrt{\sum_{i=m}^{m+N} \frac{(\Delta f_i-\Delta {f_i}^0)^2}{N}}$$, where $\Delta f_i$ is the mode spacing at $i$th mode, $\Delta {f_i}^0$ is the predicted spacing at the same mode for a disorderless line, $\overline{\Delta f}$ is an averaged mode spacing for the modes in between $m$ and $m+N$. All points at Fig.3c are obtained by averaging 30 modes around $m = 150$.

\begin{figure}[h]
\centering
\includegraphics[width=0.9\textwidth]{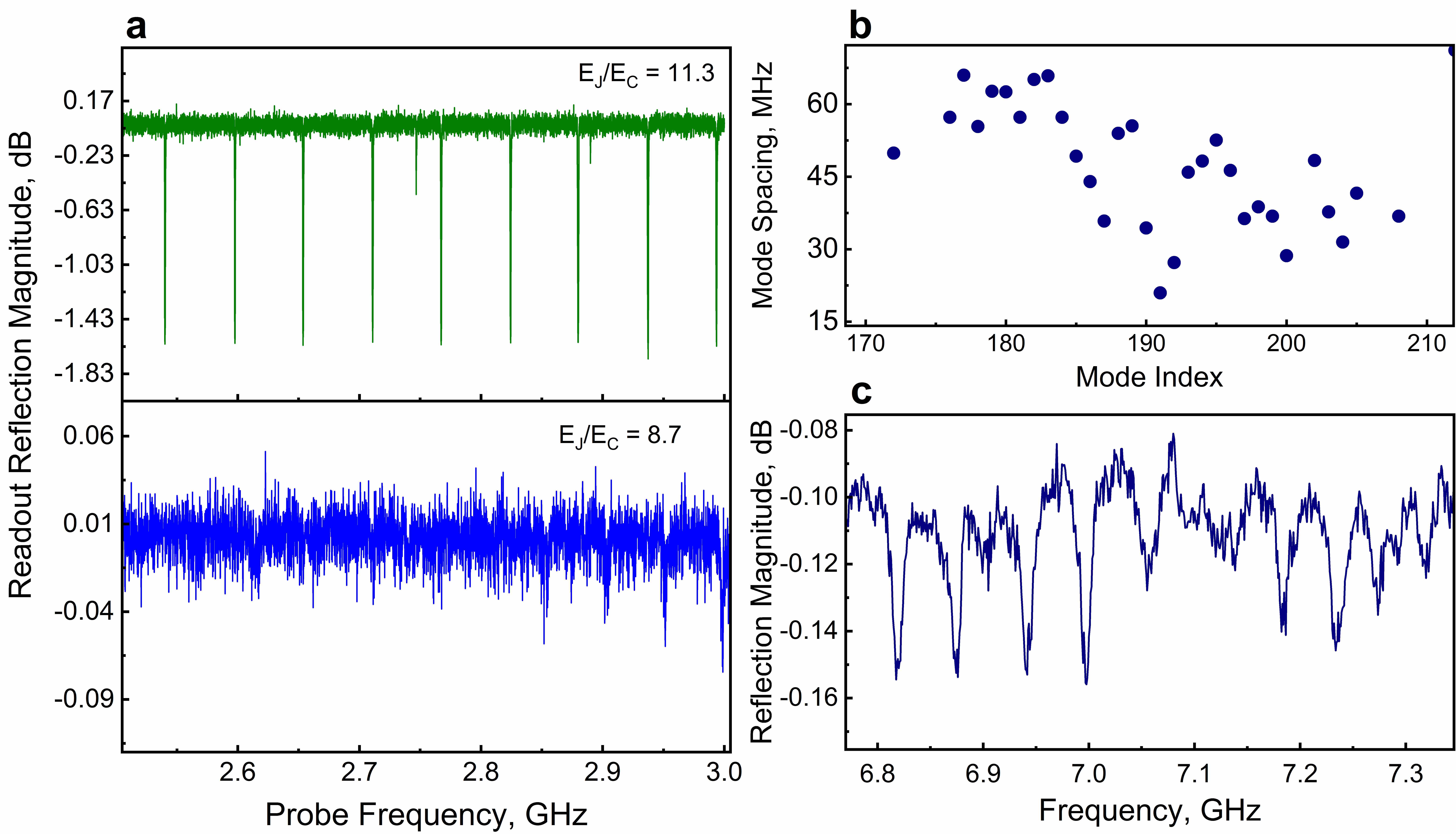}
\caption{(a) The two-tone spectroscopy of a device with tuned Ej/Ec ratio. In both plots similar read-out conditions and the same probe tone powers were used. (b) The mode spacing as a function of mode index for a weak chain. (c) One-tone spectroscopy of a weak chain showing large fluctuations in the collective mode positions.}
\end{figure}